\documentclass[11pt,twoside]{article}
\usepackage{amssymb}   
\usepackage{graphicx}
\def\ii{\'{\i}}                     % i acentuada
\def\r{\mathbb R}                   % Numeros reales
\def\n{\mathbb N}                   % Numeros naturales
\def\z{\mathbb Z}                   % Numeros enteros
\def\L{\mathbb L}                   % L del espacio de 
\def\SS{\mathbb S}                   % S del espacio de 
\def\fr{\frac}
\def\be{\begin{equation}}
\def\ee{\end{equation}}
\def\bea{\begin{eqnarray}}
\def\eea{\end{eqnarray}}
\def\b*{\begin{eqnarray*}}
\def\e*{\end{eqnarray*}}
\def\N{\hfill \rule{2.5mm}{2.5mm}}
\def\T{\tilde{{\bf g}}}

\def\U{{\cal U}}
\def\G{{\bf g}}
\def\V{\it V}
\def\n{\it n}
\def\DP{{\cal DP}}
\def\C{{\cal C}}
\def\b{\bar}
\def\g{\gamma}

\def\Z{\Theta}
\def\z{\theta}
\def\S{\Sigma}
\def\O{\Omega}
\def\L{\Lambda}
\def\p{\Psi}
\def\f{\varphi}
\def\x{\chi}
\def\P{{\it Proof :} \hspace{3mm}}
\def\X{\vec{X}}
\def\Y{\vec{Y}}
\def\k{\vec{k}}
\def\A{\mbox{A}}

\def\xiv{\vec \xi }
\def\lie{{\pounds}_{\xiv}\, }
\newtheorem{defi}{Definition}[section]
\newtheorem{theo}{Theorem}[section]
\newtheorem{coro}{Corollary}[section]
\newtheorem{prop}{Proposition}[section]
\newtheorem{lem}{Lemma}[section]

\begin{document}

\pagestyle{myheadings}
\markboth
{{\sc Causal relations and their applications }}  
{{\sc A. Garc\ii a-Parrado and J. M. M. Senovilla}}        
\vspace{1cm}

%%%%%%%%%%%%%%%%%%%%%%%%%%%%%%%%%%%%%%%%%%%%%%%%%%%%%%%%%%%%%%%%
% 
% Titulo Articulo, Autor y afiliaci\'on
% 
%%%%%%%%%%%%%%%%%%%%%%%%%%%%%%%%%%%%%%%%%%%%%%%%%%%%%%%%%%%%%%%%

\thispagestyle{empty}

\begin{center}

{\Large{\bf{Causal relations and their applications}}} 

%\medskip

\vskip1cm

{\sc Alfonso Garc\ii a-Parrado and Jos\'e M. M. Senovilla}   
\vskip0.5cm

{\it Departamento de F\ii sica Te\'orica. Universidad del Pa\ii s 
Vasco.\\
Apartado 644, 48080 Bilbao, Spain}         

\end{center}
\bigskip

%%%%%%%%%%%%%%%%%%%%%%%%%%%%%%%%%%%%%%%%%%%%%%%%%%%%%%%%%%%%%%%%
%
%   Abstract
%
%%%%%%%%%%%%%%%%%%%%%%%%%%%%%%%%%%%%%%%%%%%%%%%%%%%%%%%%%%%%%%%%
\begin{abstract}
In this work we define and study the relations between Lorentzian
manifolds given by the diffeomorphisms which map causal future 
directed vectors onto causal future directed vectors. This class of 
diffeomorphisms, called {\em proper causal relations}, contains as a 
subset 
the well-known group of conformal relations and are deeply linked to
the so-called causal tensors of Ref.\cite{S-E}. If two given 
Lorentzian manifolds are in {\em mutual} proper causal relation then
they are said to be causally isomorphic: they are indistinguishable 
from 
the causal point of view. Finally, the concept of causal 
transformation for Lorentzian manifolds is introduced and its main 
mathematical properties briefly investigated.
\end{abstract} 

%%%%%%%%%%%%%%%%%%%%%%%%%%%%%%%%%%%%%%%%%%%%%%%%%%%%%%%%%%%%%%%%
% 
%   Texto
% 
%%%%%%%%%%%%%%%%%%%%%%%%%%%%%%%%%%%%%%%%%%%%%%%%%%%%%%%%%%%%%%%%

\section{Basics on causal relations}
In this section the definitions of the basic concepts and the 
notation to be used throughout this contribution shall be presented.
Differentiable manifolds are denoted by 
italic capital letters $\V , W, U, \dots$ and, to our purposes,
all such manifolds will be connected causally orientable Lorentzian
manifolds of dimension $\n$.
The signature convention is set to $(+ -\dots -)$. $T_x(V)$ and 
$T^{*}_x(V)$ will stand respectively for the tangent and cotangent 
spaces at $x\in V$, and $T(\V)$ (resp. $T^{*}(\V)$) is the tangent 
bundle (cotangent bundle) of $\V$. Similarly 
the bundle of $j$-contravariant and $k$-covariant tensors of $V$ 
is denoted  ${\cal T}^{j}_{k}(V)$. If $\f$ is a diffeomorphism 
between $V$ and $W$, the push-forward and 
pull-back are written as $\f^{'}$ and $\f^{*}$ respectively. The 
hyperbolic structure of the Lorentzian scalar product naturally 
splits the elements of $T_x(V)$ into timelike, spacelike, and null,
and as usual we use the term {\it causal} for the vectors (or vector 
fields) which are non-spacelike. 
To fix the notation we introduce the sets:
\begin{eqnarray*}
\Z^{+}(x)&=&\{\X\in T_x(V) : \X\ \mbox{is causal future directed} 
\},\\   \Z(x)&=&\Z^{+}(x)\cup\Z^{-}(x),\hspace{1cm} 
\Z^{+}(V)=\bigcup_{x\in V}\Z^{+}(x)
\e*
with obvious definitions for $\Z^{-}(x)$, $\Z^{-}(V)$ and $\Z(V)$.   
Before we proceed, we need to introduce a further concept taken from 
\cite{S-E}.
\begin{defi}   
A tensor ${\bf T}\in {\cal T}^{0}_{r}(x)$ satisfies the 
dominant property if for every $\k_1, \dots, \k_r \in \Z^{+}(x)$
we have that ${\bf T}(\k_1, \dots, \k_r )\geq 0$.
\end{defi}
The set of all $r$-tensors with the 
dominant property at $x\in V$ will be denoted by $\DP^{+}_r(x)$ 
whereas 
$\DP^{-}_r(x)$ is the set of tensors such that $-{\bf T}\in 
\DP^{+}_r(x)$.
We put $\DP_r(x)\equiv \DP^{+}_r(x)\cup \DP^{-}_r(x)$.
All these definitions extend straightforwardly to 
the bundle ${\cal T}^{0}_{r}(V)$ and we may define the subsets 
$\DP^{+}_r(\U)$, $\DP^{-}_r(\U)$ and $\DP_{r}(\U)$ for an open subset 
$\U\subseteq V$ as follows:
\[
\DP^{\pm}_r(U)=\bigcup_{x\in \U}\DP^{\pm}_r(x),\, \, \, 
\DP_r(\U)=\DP^{+}_r(\U)\cup\DP^{-}_r(\U).
\]     
The simplest example (leaving aside $\r^+$) of causal tensors are
the causal 1-forms ($\equiv \DP_{1}(V)$) \cite{S-E}\footnote{See also
Bergqvist's and Senovilla's contributions to this volume.}, while
a general characterization of $\DP^{+}_r\equiv \DP^{+}_r(V)$ is the 
following
(see \cite{SUP} for a proof)$^1$:
\begin{prop}
${\bf T}\in \DP^{+}_r$ if and 
only if the components $T_{i_1\dots i_r}$ of ${\bf T}$ in all 
orthonormal
bases fulfill $T_{0\dots 0}\geq|T_{i_1\dots i_r}|$, $\forall i_1\dots 
i_r$,
where the $0$-index refers to the temporal component.
\label{ORTHONORMAL}
\end{prop}
%\P see \cite{SUP}.\N 

We are now ready to present our main concept, which 
tries to capture the notion of some kind of relation between the 
causal structure of $V$ and $W$.
\begin{defi}
Let $\f:V\rightarrow W$ be a global diffeomorphism between
two Lorentzian manifolds. We shall say that $W$ is 
properly causally related with $V$ by $\f$, denoted 
$V\prec_{\f}W$, if for every $\X\in\Z^{+}(V)$ we have that $\f^{'}\X$ 
belongs to $\Z^{+}(W)$. $W$ is said to be properly 
causally related with $V$, denoted simply as $V\prec W$,
if $\exists \f$ such that $V\prec_{\f}W$.
\label{PREC}
\end{defi}
{\bf Remarks}
\begin{enumerate}
\item This definition can also be given for any set $\zeta\subseteq V$
by demanding that
$(\f^{'}\X)_{\f(x)}\in \Z^{+}(\f(x))\,\,\,\, \forall \X\in\Z^{+}(x)$,
$\forall x\in\zeta$.
\item  Two diffeomorphic Lorentzian manifolds may fail to be properly
causally related as we shall show later with explicit examples.  
\end{enumerate}
\begin{defi}
Two Lorentzian manifolds $V$ and $W$ are called causally isomorphic 
if $V\prec W$ and $W\prec V$. This shall be written as $V\sim W$.
\label{EQUIV}
\end{defi}
We claim that if $V\sim W$ then their causal structure are somehow 
the same.

Let $\G$ and $\tilde{\G}$ be the Lorentzian metrics of $V$ 
and $W$ respectively. By using
\be
\tilde{\G}(\f^{'}\X,\f^{'}\Y)=\f^{*}\tilde{\G}(\X,\Y),
\label{pull-back}
\ee
we immediately realize that  
$V\prec_{\f} W$ implies that $\f^{*}\tilde{\G}\in\DP^{+}_2(V)$. 
Conversely, if $\f^{*}\T\in\DP^{+}_2(V)$ then for every 
$\X\in\Z^{+}(V)$ we have that 
$(\f^{*}\T)(\X,\X)=\T(\f^{'}\X,\f^{'}\X)\geq 0$ and hence 
$\f^{'}\X\in\Z(W)$. However, it can happen that $\Z^{+}(V)$ is
actually mapped 
to $\Z^-(W)$, and $\Z^{-}(V)$ to $\Z^+(W)$. This only means 
that $W$ {\em with the time-reversed orientation} is properly causally
related with $V$. Keeping this in mind, the
assertion $\f^{*}\T\in\DP^{+}_2(V)$ will be henceforth taken as 
equivalent to $V\prec_{\f}W$.

\section{Mathematical properties}
Let us present some mathematical properties of proper 
causal relations.
\begin{prop}    
If $V\prec_{\f}W$ then:
\begin{enumerate}
\item $\X\in\Z^{+}(V)$ is timelike $\Longrightarrow$ $\f^{'}\X\in 
\Z^{+}(W)$ is timelike.  
\item $\X\in\Z^{+}(V)$ and $\f^{'}\X\in \Z^{+}(W)$ is  null 
$\Longrightarrow$ $\X$ is null.
\end{enumerate}
\label{CAUS}
\end{prop}
\P For the first implication, if $\X\in\Z^{+}(V)$ is timelike we have,
according to equation (\ref{pull-back}), that 
$\f^{*}\tilde{\G}(\X,\X)=\T(\f^{'}\X,\f^{'}\X)$ which must be a 
strictly positive quantity as $\f^{*}\tilde{\G}\in\DP^{+}_2(W)$ 
\cite{S-E}.
For the second implication, equation (\ref{pull-back}) implies 
$0=\f^{*}\T(\X,\X)$ which is only possible if $\X$ is null since 
$\f^{*}\T\in\DP^{+}_2(V)$ and $\X\in\Z^{+}(V)$ (see again 
\cite{S-E}).\N
\begin{prop}
$V\prec_{\f}W \hspace{2mm} \Longleftrightarrow \hspace{2mm}
\f^{'}\X\in\Z^{+}(W)$ for all null $\X\in\Z^{+}(V)$.
\label{NULL}
\end{prop}
\P For the non-trivial implication, making again use of 
(\ref{pull-back}) we can write:
\[
\f^{'}\X\in\Z^{+}(W)\ \forall\X\ \mbox{null in}\ 
\Z^{+}(V)\Leftrightarrow\f^{*}\tilde{\G}(\X,\Y)\geq 0\ \ \forall\ 
\X,\Y\ \mbox{null in}\ \Z^{+}(V)
\]
which happens if and only if $\f^{*}\tilde{\G}$ is in $\DP^{+}_2(V)$
(see \cite{S-E} property 2.4).\N
\begin{prop}[Transitivity of the proper causal 
relation]\hspace{0.1cm}    \\ If $V\prec_{\f} W$ and $W\prec_{\psi} 
U$ then $V\prec_{\psi\circ\f}U$
\label{ORDER}
\end{prop}
\P  Consider any $\X\in \Z^{+}(V)$. Since 
$V\prec_{\f} W$, $\f^{'}\X\in \Z^{+}(W)$ and since $W\prec_{\psi} U$ 
we get $\psi^{'}[\f^{'}\X]\in\Z^{+}(U)$ so that 
$(\psi\circ\f)^{'}\X\in\Z^{+}(U)$ from what we conclude that 
$V\prec_{\psi\circ\f} U$.\N

Therefore, we see that the relation $\prec$ is a preorder. Notice 
that if $V\sim W$ (that is $V\prec W$ and $W\prec V$) this does not 
imply that $V=W$. Nevertheless, one can always define a partial order 
for the corresponding classes of equivalence. 

Next, we identify the part of the boundary of the null cone which
is preserved under a proper causal relation. A lemma is needed first.
Recall that $\X$ is called an ``eigenvector'' of a 2-covariant tensor 
${\bf T}$ if ${\bf T}(\cdot ,\X )=\lambda \G (\cdot ,\X )$ and 
$\lambda$ is then the corresponding eigenvalue.
\begin{lem}
If ${\bf T}\in \DP^{+}_2(x)$ and $\X\in\Z^{+}(x)$ then ${\bf 
T}(\X,\X)=0\ \Longleftrightarrow\X$ is a null eigenvector of ${\bf 
T}$.
\label{NULL-EIGEN}
\end{lem}
\P Let $\X\in\Z^{+}(x)$ and assume $0={\bf 
T}(\X,\X)=T_{ab}X^{a}X^{b}$.
Then since $T_{ab}X^{b}\in\DP^{+}_1(x)$ \cite{S-E} 
we can conclude that $X_a$ and $T_{ab}X^{b}$ must be proportional 
which results in $X^{a}$ being a null eigenvector of $T_{ab}$. The 
converse is straightforward.\N
\begin{prop}
Assume that $V\prec_{\f} W$ and $\X\in\Z^{+}(x),\ x\in V$. Then 
$\f^{'}\X$ is null at $\f(x)\in W$ if and only if $\X$ is a null 
eigenvector of 
$\f^{*}\tilde{\G}(x)$.
\label{CONE}
\end{prop}
\P Let $\X$ be in $\Z^{+}(x)$ and suppose $\f^{'}\X$ is 
null at $\f(x)$. Then, according to proposition \ref{CAUS}, $\X$ is 
also null at $x$. On the other hand we have
$0=\tilde{\G}(\f^{'}\X,\f^{'}\X)=\f^{*}\tilde{\G}(\X,\X)$
and since $\f^{*}\T|_{x}\in\DP^{+}_2(x)$, lemma \ref{NULL-EIGEN} 
implies that $\X$ is a null eigenvector of $\f^{*}\T$ at $x$.\N

The vectors which remain null under the causal relation $\f$ are 
called its {\em canonical null directions}. On the other hand, the 
null eigenvectors of ${\bf T}\in\DP^{+}_2$ can be used to 
classify this tensor, as proved in \cite{S-E}. As a result we have
\begin{prop}
If the relation $V\prec_{\f}W$ has n linearly independent canonical 
null directions then $\f^{*}\T=\lambda\G$.
\label{CONF}
\end{prop}
\P If there exist $n$ independent canonical null directions, then 
$\f^{*}\T$ has $n$ independent null eigenvectors which is only 
possible if $\f^{*}\T$ is proportional to the metric tensor $\G$ 
(\cite{S-E,SUP}.)\N

Proposition \ref{CONF} has an interesting application in the 
following 
theorem
\begin{theo}
Suppose that $V\prec_{\f}W$ and $W\prec_{\f^{-1}}V$. Then
$\f^{*}\T=\lambda\G$ and 
$(\f^{-1})^{*}\G=\fr{1}{(\f^{-1})^{*}\lambda}\T$
for some positive function $\lambda$ defined on $V$.
\label{INV}
\end{theo}
\P Under these hypotheses, using proposition \ref{CAUS}, we get the 
following intermediate results
\bea
\f^{'}\X\in \Z^{+}(W)\ \mbox{null and 
$\X\in\Z^{+}(V)$}\Longrightarrow\X\ \mbox{is null,}\nonumber\\
(\f^{-1})^{'}\Y\in\Z^{+}(V)\ \mbox{null and 
$\Y\in\Z^{+}(W)$}\Longrightarrow\Y\ \mbox{is null.}\nonumber
\eea 
Now, let $\X\in \Z^{+}(V)$ be null and 
consider the unique $\Y\in T(V)$ such that $\X=(\f^{-1})^{'}\Y$.
Then $\Y=\f^{'}\X$ and $\Y\in\Z^{+}(W)$ as $\f$ sets a proper
causal relation and $\X$ is in $\Z^{+}(V)$.
Hence, according to the second result above 
$\Y$ must be null and we conclude that every null $\X\in \Z^{+}(V)$ 
is push-forwarded to a null vector of $\Z^{+}(W)$ which in turn 
implies that $\f^{*}\T=\lambda\G$. In a similar fashion, we can prove 
that $(\f^{-1})^{*}\G=\mu\T$ and hence 
$(\f^{-1})^{*}\lambda =1/\mu$.\N
\begin{coro}
$V\prec_{\f}W$ and $W\prec_{\f^{-1}}V \,\, \Longleftrightarrow \,\, 
\f$ is a 
conformal relation.
\end{coro}

\section{Applications to causality theory}
In this section we will perform a detailed study of how two 
Lorentzian ma-\\
nifolds $V$ and $W$ such that $V\prec_{\f}W$ share 
common causal features. To begin with, we must recall the basic sets 
used in causality theory, namely $I^{\pm}(p)$ and $J^{\pm}(p)$ for 
any point $p\in V$ (these definitions can also 
be given for sets). One has $q\in J^+(p)$ (respectively 
$q\in I^+(p)$) if there exists a continuous future 
directed causal (resp.\ timelike) curve joining $p$ and $q$.
Recall also the Cauchy developments $D^{\pm}(\zeta)$ for any set
$\zeta\subseteq V$ \cite{FF,W,COND}. Another important 
concept is that of future set: $\A\subset V$ is said to be a future 
set if $I^{+}(\A)\subseteq \A$. For example $I^{+}(\zeta)$ is a 
future set for any $\zeta$.  All 
these concepts are standard in causality theory and are defined in 
many references, see for instance \cite{FF,W,COND}. 
\begin{prop}
If $V\prec_{\f} W$ then, for every set 
$\zeta\subseteq V$, we have $\f(I^{\pm}(\zeta))\subseteq 
I^{\pm}(\f(\zeta))$ and $\f(J^{\pm}(\zeta))\subseteq 
J^{\pm}(\f(\zeta))$.
\label{SET}
\end{prop}
\P It is enough to prove it for a single point $p\in V$ and then
getting the result for every $\zeta$ by considering it as 
the union of its points. For the first relation, let $y$ be in 
$\f(I^{+}(p))$ arbitrary and take $x\in I^{+}(p)$ such that 
$\f(x)=y$. 
Choose a future-directed timelike curve $\g$ joining $p$ 
and $x$. From proposition \ref{CAUS}, $\f(\g)$ is then a 
future-directed timelike curve joining $\f(p)$ and $y$, so that
$y\in I^{+}(\f(p))$. The second assertion is proven in a similar way 
using again proposition \ref{CAUS}.
The proof for the past sets is analogous.\N

The converse of this proposition does not hold in general unless we impose some causality conditions on the spacetime.
\begin{defi}
A Lorentzian manifold $V$ is said to be distinguishing if for every neighbourhood $U_p$ of $p\in V$ there exist another neighbourhood $B_p\subset U_p$ containing $p$ which intersects every causal curve meeting $p$ in a connected set. 
\label{DISTINGUISHING}
\end{defi}
  We need some concepts of standard causality theory. For any $p\in V$ one can introduce normal coordinates in a neighbourhood
${\cal N}_p$ of $p$ (see, e.g. \cite{COND}). Then the exponential map
provides a diffeomorphism $\exp:{\cal O}\subset T_p(V)\rightarrow 
{\cal N}_p$
where ${\cal O}$ is an open neighbourhood of $\vec{0}\in T_p(V)$.
The interior of the future (past) light cone of $p$ is defined by
$C^{\pm}_p=\exp(\mbox{int}(\Z^{\pm}(p))\cap {\cal O})$, and obviously
$C^{\pm}_p\subseteq I^{\pm}(p)$ \cite{COND}. Other important issue deals with the chronology relation $<\!\!<$ between two points. We have $p<\!\!<q$ if there exist a future timelike curve joining $p$ and $q$. See \cite{KRONHEIMER} for an axiomatic study of this relation. 
\begin{prop}
Let $\g$ be a piecewise continuous curve of a distinguishing Lorentzian manifold $(V,\G)$. Then, $\g$ is total with respect to $<\!\!<$ if and only if $\g$ is timelike.
\end{prop}
\P Clearly if $\g$ is timelike then $\g$ must be a total set for the relation $<\!\!<$ (this is true for every spacetime). For the converse consider a curve $\g$ which is total with respect to $<\!\!<$ and let $q\in\g$ be an arbitrary point of the curve. If we take a normal neighbourhood of $q$, ${\cal N}_q$ then we can find a neighbourhood $U_q$ of $q$ which is intersected in a connected set by every causal curve meeting $q$. Now, if we pick up a point $z\in\g\cap U_q$ we have that either $q<\!\!<z$ or $z<\!\!<q$. Assuming the former we deduce that there exists a timelike curve $\tilde{\g}$ joining $q$ and $z$ which implies that $\tilde{\g}\cap U_q$ is a connected set. This property together with the distinguishability of $V$ implies that $\tilde{\g}$ must be a subset of $U_q$ and hence $\tilde{\g}\subset {\cal N}_q$ from what we conclude that $\tilde{\g}\subset C_p$ (\cite{COND}) and hence $z\in C_p$ $\forall z\in\g\cap U_q$ which is only possible if $\g\cap U_q$ is timelike. By covering $\g$ with sets of the form $\g\cap U_q$, $q\in \g$ we arrive at the desired result.\N
\begin{prop}
Let $\f:V\rightarrow W$ be a diffeomorphism with the property 
$\f(I^{+}(p))\subseteq I^{+}(\f(p))\ \forall p\in V$. Then if $W$ is distinguishing, $\f$ is a proper causal relation. A similar result holds replacing $I^{+}$ by $I^{-}$.
\label{CHRONOLOGICAL}
\end{prop}
\P From the statement of this proposition is clear that $\forall$ $p,q$ of $V$ such that $p<\!\!<q$ then $\f(p)<\!\!<\f(q)$. Therefore every timelike curve $\g$ of $V$ is mapped onto a continuous curve in $W$ total with respect to $<\!\!<$ and hence timelike due to the distiguishability of $W$. Furthermore if the curve $\g$ is future directed then $\f(\g)$ must be also future directed since $<\!\!<$ is preserved which is only possible if every timelike future-pointing vector is mapped onto a future-pointing timelike vector. As a consequence, if $\k$ is 
a null vector, $\f^{'}\k$ must be a causal vector (to see it just 
construct a sequence of timelike future directed vectors converging 
to $\k$) which proves that $\f$ is a proper causal relation.\N

The results for the Cauchy developments are the following:
\begin{prop}
If $V\prec_{\f} W$ then 
$D^{\pm}(\f(\zeta))\subseteq \f(D^{\pm}(\zeta)),
\,\, \forall \zeta\subseteq V$.
\label{CAUCH}
\end{prop}
\P It is enough to prove the future case. Let $y\in D^{+}(\f(\zeta))$
arbitrary and consider any causal past directed curve 
$\g^{-}_{\f^{-1}(y)}\subset V$ containing $\f^{-1}(y)$. Since the 
image curve by $\f$ of $\g^{-}_{\f^{-1}(y)}$ is a causal curve 
passing through $y$, ergo meeting $\f(\zeta)$, we have that 
$\g^{-}_{\f^{-1}(y)}$ must meet $\zeta$ from what we conclude that 
$y\in\f(D^{+}(\zeta))$ due to the arbitrariness of 
$\g^{-}_{\f^{-1}(y)}$.\N
\begin{coro}
If $\S\subset W$ is a Cauchy hypersurface then $\f^{-1}(\S)$ 
is also a Cauchy hypersurface of $V$.
\label{HYP}
\end{coro}
\P If $\S$ is a Cauchy hypersurface then $D(\S)=W$, and from 
proposition \ref{CAUCH} $D(\S)\subseteq\f(D(\f^{-1}(\S)))$.
Since $\f$ is a diffeomorphism the result follows.\N

One can prove the impossibility of the existence
of proper causal relations sometimes. For instance, from the previous 
corollary we deduce that $V\prec W$ is impossible if $W$ is 
globally hyperbolic but $V$ is not. Other impossibilities arise as 
follows. Let us recall that, for any inextendible causal curve $\g$, 
the 
boundaries $\partial I^{\pm}(\g)$ of its chronological future and 
past 
are usually called its future and past event horizons, sometimes also 
called particle horizons \cite{FF,W,COND}. Of course these sets can 
be empty (then one says that $\g$ has no horizon).
\begin{prop}
Suppose that every inextendible causal future directed curve in $W$ 
has a non-empty $\partial I^{-}(\g)$ ($\partial I^{+}(\g)$).
Then any V such that $V\prec W$ cannot have 
inextendible causal curves without past (future) event horizons.
\label{HOR}
\end{prop}
\P If there were a future-directed curve $\g$ in $V$ with
$\partial I^{-}(\g)=\emptyset$, $I^{-}(\g)$ would be the whole of 
$V$. But 
according to proposition \ref{SET} $\f(I^{-}(\g))\subseteq 
I^{-}(\f(\g))$
from what we would conclude that $I^{-}(\f(\g))=W$ against 
the assumption. \N

The class of future (or past) sets  
characterize the proper causal relations for distinguishing spacetimes as it is going to be shown next (every statement for future objects has a counterpart for the past). 
\begin{lem}
If $\A$ is a future set then $p\in\overline{\A}\Longleftrightarrow 
I^{+}(p)\subseteq\A$.
\label{CLOSURE}
\end{lem}
\P Suppose $I^{+}(p)\subseteq A$. Then since $C_p^{+}\subseteq 
I^{+}(p)$
and $p\in \overline{C_p^{+}}$ we have that $U_p\cap 
C_p^{+}\neq\emptyset$ for every neighbourhood $U_p$ of $p$ which in 
turn implies that $U_p\cap\A\neq\emptyset$ and hence 
$p\in\overline{\A}$. Conversely, let $p$ be any point of 
$\overline{\A}$ then $I^+(p)\subseteq 
I^+(\overline{A})=I^+(A)\subseteq A$.\N
\begin{theo}
Suppose that $(W,\T)$ is a distinguishing spacetime. Then a diffeomorphism $\f:(V,\G)\rightarrow (W,\T)$ is a proper causal relation if 
and only if $\f^{-1}(\A)$ is a future set for every future set.   
$\A\subseteq W$.
\label{KEY}
\end{theo}
\P Suppose $\A\subseteq W$ is a future set, $V\prec_{\f} W$ and take 
$\f^{-1}(\A)\subseteq V$. Proposition \ref{SET} implies 
$\f(I^{+}(\f^{-1}(\A)))\subseteq 
I^{+}(\f(\f^{-1}(\A)))=I^{+}(\A)\subseteq\A$ which shows that 
$I^{+}(\f^{-1}(\A))\subseteq \f^{-1}(\A)$. Conversely, for any 
$p\in V$ take the future set $I^{+}(\f(p))$ and consider the future 
set 
$\f^{-1}(I^{+}(\f(p)))$. As $\f(p)\in\overline{I^{+}(\f(p))}$ then 
$p\in\overline{\f^{-1}(I^{+}(\f(p)))}$ and according to 
lemma \ref{CLOSURE} $I^{+}(p)\subseteq\f^{-1}(I^{+}(\f(p)))$
so that $\f(I^{+}(p))\subseteq I^{+}(\f(p))$.
Since this holds for every $p\in V$ and $W$ is distinguishing, proposition 
\ref{CHRONOLOGICAL} ensures that $\f$ is a proper causal relation.\N

This theorem has important consequences.
\begin{prop}
If $V\sim W$ and both manifolds are distinguishing, then there is a one-to-one correspondence between the 
future (and past) sets of $V$ and $W$. 
\label{CONSERV}
\end{prop}
\vspace{-0.5cm}
\P If $V\sim W$ then $V\prec_{\f}W$ and $W_{\prec\Psi}V$ for some 
diffeomorphisms $\f$ and $\Psi$. By denoting with ${\cal F}_V$ and 
${\cal F}_W$ the set of future sets of $V$ and $W$ respectively, we 
have that $\f^{-1}({\cal F}_W)\subseteq{\cal F}_V$ and 
$\Psi^{-1}({\cal F}_V)\subseteq{\cal F}_W$, due to theorem \ref{KEY}.
Since both $\f$ and $\Psi$ are bijective maps we conclude that
${\cal F}_V$ is in one-to-one correspondence with a subset of ${\cal 
F}_W$
and vice versa which, according to the equivalence theorem
of Bernstein \cite{BERN}, implies that ${\cal F}_V$ is in one-to-one
correspondence with ${\cal F}_W$.\N

\section{Causal transformations}
In this section we will see how the concepts above generalize, in a 
natural way, the group of conformal transformations in a Lorentzian 
manifold $V$. 
\begin{defi}
A transformation $\f:V\longrightarrow V$ is called 
{\em causal} if $V\prec_{\f}V$.
\label{GG}
\end{defi}
The set of causal transformations of $V$ will be denoted by $\C(V)$.
This is a subset of the group of transformations of $V$ which is 
closed
under the composition of diffeomorphisms, due to proposition 
\ref{ORDER},
and contains the identity map.
This algebraic structure is well-known, see e.g. \cite{SEMIGROUP}, 
and called subsemigroup with identity or submonoid.
Thus, $\C(V)$ is a {\em submonoid} of the group of diffeomorphisms of 
$V$.
Nonetheless, $\C(V)$ usually fails to be a group. In fact we have,
\begin{prop}
Every subgroup of causal transformations is a group of conformal 
transformations.
\label{GROUP}
\end{prop}
\P Let $G\subseteq \C(V)$ be a subgroup of causal transformations and 
consider any $\f\in G$, so that both $\f$ and $\f^{-1}$ 
are causal transformations. Then $\f$ is necessarily a conformal
transformation as follows from Theorem \ref{INV}.\N

From standard results, see \cite{SEMIGROUP},
we know that $\C(V)\cap\C(V)^{-1}$ is just the 
group of conformal transformations of $V$ and there is 
no other subgroup of $\C(V)$ which contains $\C(V)\cap\C(V)^{-1}$.
The causal transformations which are not conformal transformations 
are called proper causal transformations. 

It is now a natural question whether one can define infinitesimal 
generators of one-parameter families of causal transformations which 
generalize the ``conformal Killing vectors'', and in which sense.
Notice, however, that if $\{\f_{s}\}_{s\in \r}$ is a one-parameter 
group of causal transformations, from the previous results
the only possibility is that $\{\f_{s}\}$ be in fact a group of
conformal motions. On the 
other hand, things are more subtle if there are no conformal 
transformations in the family $\{\f_s\}$ other than the identity, in 
which case it is easy to see that the `best' one can accomplish is 
that either $G^{+}\equiv \{\f_{s}\}_{s\in \r^{+}}$ or
$G^{-}\equiv \{\f_{s}\}_{s\in \r^{-}}$ is in $\C(V)$. If this 
happens one talks about maximal one-parameter submonoids of
proper causal transformations. Of course, it is also possible to
define local one-parameter submonoids of 
causal transformations $\{\f_s\}_{s\in I}$ for some interval 
$I=(-\epsilon,\epsilon)$ of the real line assuming that
$\{\f_s\}_{s\in (0,\epsilon)}$ consists of proper 
causal transformations. In any of these cases, we can define
the infinitesimal generator of $\{\f_s\}$ as the vector field 
$\xiv=d\f_{s}/ds|_{s=0}$. Given that $\f_{s}^{*}\G\in\DP_{2}$ for 
all $s\geq 0$ (or for all non-positive $s$), one can somehow control 
the 
Lie derivative of $\G$ with respect to $\xiv$. For instance, it is 
easy 
to prove that $\lie\G (\k,\k )\geq 0$ (or $\leq 0$) for all null 
$\k$, 
clearly generalizing the case of conformal Killing vectors. An 
explicit example of this will be shown in the next section.

\section{Examples}
{\bf Example 1 Einstein static universe and de Sitter spacetime.} 
Let us take $V$ as the Einstein static universe \cite{FF} and $W=\SS$ 
as 
de Sitter spacetime. In both cases the manifold is $\r\times 
S^{3}$ and hence they are diffeomorphic. By proposition \ref{HOR} we 
know that $V\not\prec W$ because every causal curve in de Sitter 
spacetime possesses event horizons. However, the proper causal 
relation in the opposite way does hold as can be shown by 
constructing it explicitly. The line element of each spacetime is 
(with the notation $d\O^2=d\z^2+\sin^2\z d\phi^2$):
\bea
V:& & ds^{2}=dt^{2}-a^2(d\x^{2}+\sin^{2}\x d\O^{2})\ \nonumber\\
W:& & 
d\tilde{s}^{2}=d\bar{t}^{2}-\alpha^{2}\cosh^{2}(\bar{t}/\alpha)
(d\bar{\x}^{2}+\sin^{2}\bar{\x} 
d\bar{\O}^{2})\ \nonumber
\eea
where $\x,\z,\phi$ (and their barred versions) are standard 
coordinates in $S^{3}$ and $a,\alpha$ are constants. The 
diffeomorphism $\p:W\rightarrow V$ is chosen as
$\{t=b\bar{t}, \x={\bar\x}, \z={\bar\z}, \phi={\bar\phi}\}$
for a constant $b$. One can readily get $\p^{*}\G$ 
\begin{eqnarray*}
(\p^{*}\G)_{ab}dx^{a}dx^{b}=
b^{2}d\bar{t}^2-a^2(d\bar{\x}^{2}+\sin^{2}\bar{\x} d\bar{\O}^{2})
\end{eqnarray*}
which on using proposition \ref{ORTHONORMAL} shows that 
$\p^{*}\G\in\DP^{+}_{2}(W)$ if $b^{2}\geq a^2/\alpha^{2}$ and 
therefore $\p$ is proper causal relation for those $b$.\\

\hspace{-0.7cm}{\bf Example 2}
Consider the following spacetimes:
$\L_{a}$ is the region of Lorentz-Minkowski spacetime with $R>a> 0$ in
spherical coordinates $\{T,R,\Theta,\Phi\}$; $W_c$ is the outer 
region of
Schwarzschild spacetime with $r>c\geq 2M$ in Schwarzschild 
coordinates 
$\{t,r,\z ,\phi\}$. Define the diffeomorphism $\f:\L_a\rightarrow 
W_c$ given by 
$\{t=b T, r=R-a+c, \z = \Theta, \phi=\Phi\}$ for an appropriate 
positive
constant $b$, so that we have
\[  
(\f^{*}\T)_{ab}dx^adx^b=
b^{2}(1-\fr{2M}{R-a+c})dT^{2}-
\fr{dR^{2}}{1-\fr{2M}{R-a+c}}-(R-a+c)^{2}d\O^{2}.
\]
By choosing $b$ and $a$ one can achieve 
$\f^{*}\T\in \DP^{+}_2(V_{a})$
{\em whenever} $c>2M$, while for $c=2M$ $\f$ fails to be a proper 
causal
relation. Actually $\L_{a}\not\prec W_{2M}$ due to corollary \ref{HYP}
as $W_{2M}$ is globally hyperbolic but $\L_{a}$ is not.

Take now the diffeomorphism
$\Psi :W_c\rightarrow \L_{a}$ defined by 
$\{T=t,R=r,\Theta=\z,\Phi=\phi\}$, so that $\Psi^{*}\G$ reads 
$(\Psi^{*}\G)_{ab}dx^adx^b=dt^2-dr^2-r^2d{\O}^2$
from where we immediately deduce that 
$\Psi^{*}\G\in\DP^{+}_2(V)$ for every $c\geq 2M$ as long as $a\geq 
2M$. We have thus proved that $W_c\sim \L_{c}$ if 
$c>2M$, but not for $c=2M$. This is quite interesting  
and clearly related to the null character of the event horizon 
$r=2M$ in Schwarzschild's spacetime.\\

\hspace{-0.7cm}{\bf Example 3 (Friedman cosmological models with 
$p=\g\rho$.)}
Let us take as $(W,\T)$ the flat Friedman-Robertson-Walker (FRW) 
spacetimes in standard FRW coordinates $\{t,\x,\z,\phi\}$ with line 
element 
given by
\begin{eqnarray*}
ds^{2}=dt^{2}-a^{2}(t)(d\x^{2}+\x^{2}d\O^{2})
\end{eqnarray*}
and assume that the source of Einstein's equations is a perfect fluid 
with equation of state given by $p=\g\rho$ ($p=$ pressure, $\rho=$ 
density, $\g\in(-1,1)$ constant). Then the scale factor is 
$a(t)=Ct^{\fr{2}{3(1+\g)}}$ with constant $C$. By straightforward 
calculations, it can be proven the following causal equivalences:  
\bea 
W\sim \L_0\  \mbox{for $\g=-1/3$ where $\L_0$ is the whole Minkowski 
spacetime }\nonumber\\
W\sim V\  \mbox{for $\g\neq -1/3$ where $(V,\G)$ is the steady 
state part of $\SS$, \cite{FF}.}\nonumber
\eea
These causal equivalences are rather intuitive if we have a look at 
the Penrose diagram of each spacetime (figure \ref{STEADY}).
\begin{figure}[h]
\begin{center}
\includegraphics[width=\textwidth]{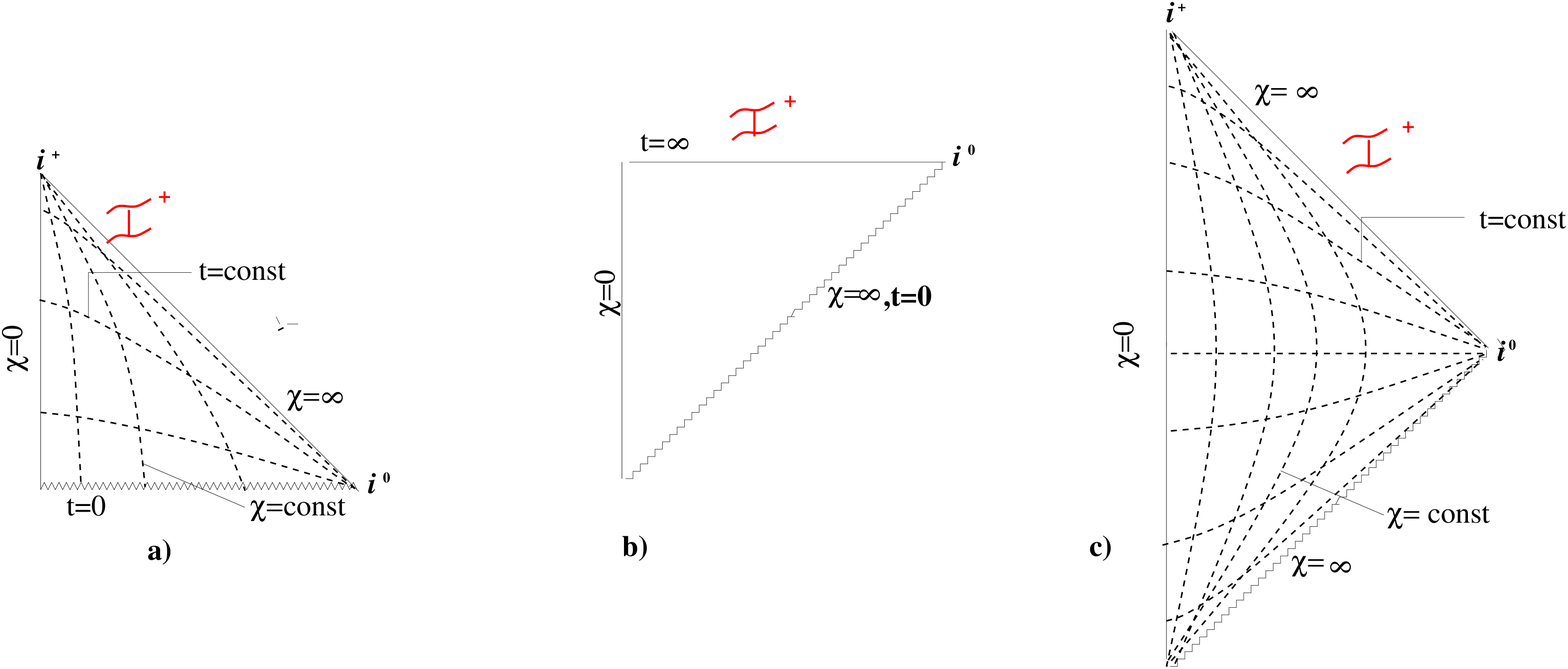}
\end{center}
\caption[]{Penrose's diagrams of FRW spacetimes for a) $-1/3<\g<1$ , 
b) $-1<\g<-1/3$ and c) $\g=1/3$. Notice the similar 
shape of diagram c) with that of $\L$, and of the
steady state part of $\SS$ with a) and b) \cite{FF}.}
\label{STEADY}
\end{figure} 

\vspace{0.3cm}

\noindent
{\bf Example 4 (Vaidya's Spacetime.)} Let us show finally 
an example of a submonoid of causal transformations.
Consider the Vaidya spacetime whose line element is \cite{V}
\[
ds^2=\left(1-\fr{2M(t)}{r}\right)dt^2-2dtdr-r^2d\O^2,\ \ 
-\infty<t<\infty,\ 0<r<\infty
\]   
where $t$ is a null coordinate (that is, $dt$ is a null 1-form), and 
$M(t)$ is a non-increasing function of $t$ interpreted as the mass.
Take the diffeomorphisms $\f_s:t\mapsto t+s$. Then $\f^{*}_s\G$ 
can be cast in the form
\[
\f^{*}_s\G=\G-\fr{2}{r}(M(t+s)-M(t))dt\otimes dt.
\]
Hence, $\f_{s}^{*}\G\in\DP^{+}_2(V)$ iff 
$M(t+s)-M(t)\leq 0$, which implies that $\{\f_s\}_{s\geq 0}$ are
causal transformations, so that $\{\f_s\}_{s\in\r}$ is a maximal 
submonoid of causal transformations. The differential equation for 
the infinitesimal generator $\vec{\xi}=\partial /\partial t$
of this submonoid is easily calculated and reads
\[
\pounds(\vec{\xi})\G=-\fr{2}{r}\dot{M}(t)dt\otimes dt \, .
\]
This is a particular case of a proper Kerr-Schild vector field, 
recently 
studied in \cite{KERR-SCHILD}. Notice that Schwarzschild spacetime is 
included for the case $M=$const., in which case $\xiv$ is a Killing 
vector. This may lead to a natural generalization of symmetries.
\section{Conclusions}
In this work a new relation between Lorentzian manifolds which keeps the causal character of causal vectors has been put forward. With the aid of this relation, we have introduced the concepts of causal relation and causal isomorphism of Lorentzian manifolds which allow us to establish rigorously when two given Lorentzian manifolds are causally indistinguishable regardless their metric properties. This tools could be also useful in order to find out the global causal structure of a given spacetime by just putting it in causal equivalence with another known spacetime.

Finally a new transformation for Lorentzian manifolds, called causal transformation has been defined. These transformations are a natural generalization of the group of conformal transformations and their actual relevance is one of our main lines of future research.
\section{Acknowledgements}
This research has been carried out under the research project UPV 
172.310-G02/99 of the University of the Basque Country.

%%%%%%%%%%%%%%%%%%%%%%%%%%%%%%%%%%%%%%%%%%%%%%%%%%%%%%%%%%%%%%%%

\end{document}